\documentclass[twocolumn,secnumarabic,amssymb,nobibnotes,aps,superscriptaddress,pra]{revtex4-1}
\pdfoutput=1 

\usepackage[dvips]{graphicx}
\usepackage{bm}
\usepackage{latexsym} 
\usepackage{amsmath}
\usepackage{amssymb}
\usepackage{color}

\newcommand{\ui}{{\rm i}}

\newcommand{\bmH}{{\bm H}}

\newcommand{\bmr}{{\bm r}}
\newcommand{\bmp}{{\bm p}} 

\newcommand{\bmq}{{\bm q}}

\newcommand{\bmu}{{\bm u}}

\newcommand{\bmS}{{\bm S}}

\newcommand{\bmsig}{{\bm \sigma}}

\newcommand{\bra}{\langle}
\newcommand{\ket}{\rangle}
\newcommand{\kB}{k_{\rm B}}

\makeatletter
\renewcommand*{\p@subsection}{}
\renewcommand*{\p@subsubsection}{}
\makeatother

\begin{document}

\title{
  Spin pumping into a spin glass material 
}

\author{Yusei Fujimoto} 
\affiliation{Department of Physics, Okayama University, Okayama 700-8530, Japan}
\author{Masanori Ichioka}
\affiliation{Research Institute for Interdisciplinary Science, Okayama University, Okayama 700-8530, Japan}
\affiliation{Department of Physics, Okayama University, Okayama 700-8530, Japan}
\author{Hiroto Adachi}
\affiliation{Research Institute for Interdisciplinary Science, Okayama University, Okayama 700-8530, Japan}
\affiliation{Department of Physics, Okayama University, Okayama 700-8530, Japan}
\date{\today}

\begin{abstract}
  Spin pumping is a recently established means for generating a pure spin current, whereby spins are pumped from a magnet into the adjacent target material under the ferromagnetic resonance condition. We theoretically investigate the spin pumping from an insulating ferromagnet into spin glass materials. Combining a dynamic theory of spin glasses with the linear-response formulation of the spin pumping, we calculate temperature dependence of the spin pumping near the spin glass transition. The analysis predicts that a characteristic peak appears in the spin pumping signal, reflecting that the spin fluctuations slow down upon the onset of spin freezing. 
\end{abstract} 



\maketitle

\section{Introduction \label{Sec:I}}
Spin current is a flow of spin angular momentum~\cite{Maekawa13}. Over the last two decades, great progress has been made in generating, manipulating, and detecting the spin current~\cite{Zutic04,SpinCurrent}. With regard to the spin current generation, as nicely reviewed in Ref.~\cite{Zutic-Dery11} the spin pumping is now established as a charge-free and versatile means~\cite{Urban01,Mizukami02,Tserkovnyak02,Saitoh06}. In this method a pure spin current, which is unaccompanied by a charge current, is pumped from a ferromagnet into the adjacent spin sink material by a stimulus of microwaves satisfying the ferromagnetic resonance (FMR) condition. Thanks to the advent of the spin pumping technique, the spin current physics has so far been investigated in a variety of spin sink materials, ranging from nonmagnetic metals~\cite{Mosendz10,Sandweg11,Czeschka11,Castel12,Hahn13}, semiconductors~\cite{Ando11,Rojas13,Shikoh13}, magnetic metals~\cite{Hyde14,Mendes14,Saglam16}, insulators~\cite{Wang14}, to more exotic systems such as graphene~\cite{Patra12,Tang13,Dushenko16}, transition metal dichalcogenides~\cite{Mendes18}, organic materials~\cite{Ando13,Qiu15}, and strongly spin-orbit coupled materials~\cite{Shiomi14, Sanchez13}. 

Recently, the playground of the spin current physics has been extended to disordered magnets or the so-called spin glass (SG) materials~\cite{Niimi15,Wesenberg17}. The SGs are characterized by a freezing of random spins~\cite{ill-cond}, and its nature has long been studied both experimentally and theoretically~\cite{Binder86}. However, despite its long history of research, the interplay of spin current and the SG ordering has not yet been well examined. Thus, it is quite natural to ask what happens if we inject a pure spin current into a SG material by the spin pumping. Experimentally, the spin pumping into a SG material was reported in 2011 using a Ag$_{90}$Mn$_{10}$/Ni$_{81}$Fe$_{19}$ bilayer~\cite{Iguchi11}. To the best of our knowledge, however, no theoretical work on the spin pumping into SG materials can be found in the literature. Therefore, developing a theory of spin pumping into the SG material is highly desirable. 

In this paper, we theoretically investigate the spin pumping into SG materials. Although a metallic magnet Ni$_{81}$Fe$_{19}$ was used as the spin injecting magnet in the previous experiment~\cite{Iguchi11}, we consider here an insulating ferromagnet such as yttrium iron garnet for the spin injector, since it makes the spin pumping signal more visible. Our strategy to calculate the spin pumping into SG materials is as follows. First, we use a linear-response approach to the spin pumping~\cite{Ohnuma14,Ohnuma15}. The notion derived from the linear response approach, that the spin pumping is intimately related to the dynamic spin susceptibility of the spin sink layer, has successfully been applied to the spin pumping into a ferromagnet~\cite{Khodadadi17}, antiferromagnets~\cite{Frangou16,Qiu16}, and recently it was also applied to the spin pumping into superconductors~\cite{Inoue17,Yao18,Umeda18}. Thus, we relate the spin pumping signal to the dynamic spin susceptibility of the SG layer. Next, we calculate the susceptibility of the SG layer by employing a dynamic theory of SGs~\cite{Sompolinsky82,Sompolinsky82b}. Not only that this theory is known to be an alternative formulation of the static replica theory~\cite{Edwards75,Sherrington75,Parisi79,Parisi80}, but also that the dynamic theory is more suitable to discuss the dynamic quantity such as the dynamic spin susceptibility~\cite{Fischer84a}.

In the literature, the dynamic spin susceptibility near the SG transition was calculated~\cite{Fischer84a}, but the result was limited to the Ising case and to an extremely low-frequency regime less than $10$~KHz, which is out of the FMR condition. In the present paper, we extend the susceptibility calculation of Ref.~\cite{Fischer84a} to the Heisenberg case and GHz frequency regime that is relevant to spin pumping experiments, and calculate temperature dependence of the spin pumping into SG materials. With this, we show that a characteristic peak structure appears near the SG transition, which is a consequence of the slowing down of spin fluctuations that is concomitant with the spin freezing of the system. 

The plan of this paper is as follows. In the next section, we introduce our microscopic model, and relate the spin pumping with the dynamic spin susceptibility of impurity spins. In Sec.~\ref{Sec:III}, on the basis of the dynamic theory of SGs, we explain how to calculate the dynamic spin susceptibility of impurity spins. In Sec.~\ref{Sec:IV}, the spin pumping signal into a SG material is calculated as a function of temperature. Finally, in Sec.~\ref{Sec:V} we discuss and summarize our results.

\section{Model \label{Sec:II}}
\begin{figure}[t] 
  \begin{center}
    \includegraphics[width=8.0cm]{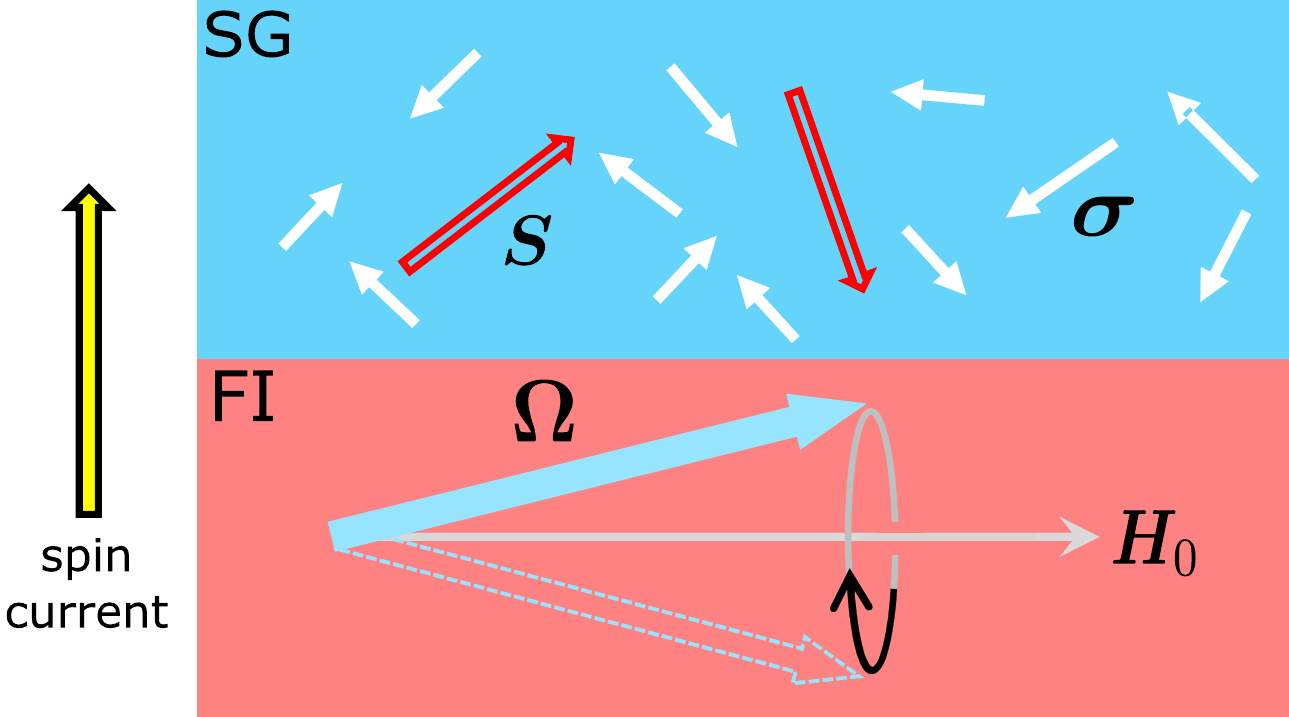}
  \end{center}
  \caption{Schematics of the system considered in this paper, where the bilayer is composed of a SG material and a ferromagnetic insulator (FI). Here, ${\bm \sigma}$ and $\bmS$ are, respectively, the conduction-electron spin and the impurity spin in the SG layer, and ${\bm \Omega}$ is the localized spin in the FI layer. A spin current with a helicity opposite to ${\bm \Omega}$ flows from the FI layer to the SG layer.}
  \label{fig:schematic01}
\end{figure}

We consider a bilayer composed of a ferromagnetic insulator (FI) and a SG material, as shown in Fig.~\ref{fig:schematic01}. More concretely, we may think of yttrium iron garnet (YIG) for the FI layer and Mn-doped Cu (Cu:Mn) for the SG layer. We assume that a static magnetic field $\bmH_0= H_0 {\bf \hat{z}}$ is applied to the FI/SG bilayer in the lateral direction, and that the anisotropy field is much smaller than $H_0$ such that it can be discarded.

We start from the following Hamiltonian: 
\begin{eqnarray}
  {\cal H} &=& {\cal H}_{\rm SG} + {\cal H}_{\rm FI-SG},
  \label{eq:H_tot01}
\end{eqnarray}
where the first term,
\begin{equation}
  {\cal H}_{\rm SG} = \sum_{\bmp} \xi_\bmp c^\dag_{\bmp} c_{\bmp}
  + J_{\rm eS} \sum_{\bmr_a} \bmsig (\bmr_a) \cdot \bmS(\bmr_a), 
  \label{eq:H_SG01} 
\end{equation}
describes the SG layer~\cite{Larkin71,Fischer79}. Here, the first term on the right-hand side describes the conduction electron kinetic energy, and the second term the coupling between the conduction electron spin and magnetic impurity at an impurity position $\bmr_a$, where $J_{\rm eS}$ is the $sd$-type exchange coupling. Here, $c^\dag_\bmp= (c^\dag_{\bmp, \uparrow}, c^\dag_{\bmp, \downarrow})$ is the electron creation operator for spin projection $\uparrow$ and $\downarrow$, $\bmS$ is an impurity spin, $\bmsig (\bmr) = c^\dag(\bmr) {\bf \hat{\bmsig}} c(\bmr) $ is the spin density operator with ${\bf \hat{\bmsig}}$ being the Pauli matrices, and $c(\bmr)= N_{\rm SG}^{-1/2} \sum_\bmp c_\bmp e^{\ui \bmp \cdot \bmr}$ with $N_{\rm SG}$ being the number of lattice sites at the SG layer.

The second term of Eq.~(\ref{eq:H_tot01}), 
\begin{eqnarray}
  {\cal H}_{\rm FI-SG} &=& J_{\rm int} \sum_{\bmr_{\rm int}} \bmsig(\bmr_{\rm int}) \cdot {\bm \Omega}(\bmr_{\rm int}), 
  \label{eq:H_FISG01} 
\end{eqnarray}
describes the interaction between the FI and SG layers. Here, $J_{\rm int}$ is the interfacial $sd$ coupling between the conduction electron spins in the SG and the localized spins in the FI, where $\bmr_{\rm int}$ is a position at the FI/SG interface. 

In order to investigate the spin pumping in the present system, we use the linear-response formulation of the spin pumping~\cite{Ohnuma14,Ohnuma15}. We consider the situation where an external microwave with the angular frequency $\omega_{\rm ac}$ is applied to the FI/SG bilayer that drives the FMR of the FI side. The linear-response formulation uses the following magnon language. In the absence of the adjacent SG layer, the uniform-mode (Kittel mode) magnon has an intrinsic damping rate $\alpha_0 \omega_{\rm ac}$, where $\alpha_0$ is the intrinsic Gilbert damping constant. In the presence of the SG layer, since the spin-relaxation rate due to the SG layer is additive and hence an additional spin dissipation channel opens, there arises an additional magnon damping rate. Therefore, the total Gilbert damping constant $\alpha$ for the bilayer is given by 
\begin{equation}
  \alpha= \alpha_0 + \delta \alpha, 
\end{equation}
where $\delta \alpha$ is the additional Gilbert damping constant. The relationship between this additional Gilbert damping constant and the spin current $I_s$ pumped into the SG layer with $z$-axis polarization is given by~\cite{Ohnuma14} 
  \begin{equation}
    I_s = \delta \alpha \frac{\gamma \hbar}{ M_s V}
    \frac{\omega_{\rm ac} (\gamma h_{\rm ac})^2}{(\gamma H_0 - \omega_{\rm ac})^2 + (\alpha_0 \omega_{\rm ac})^2}, 
  \end{equation}
  where $\gamma$ is the gyromagnetic ratio, $M_s$ the saturation magnetization, $V$ the volume of the magnet, and $h_{\rm ac}$ is the amplitude of the external microwave.

\begin{figure}[t]
  \begin{center}
    \includegraphics[width=8.0cm]{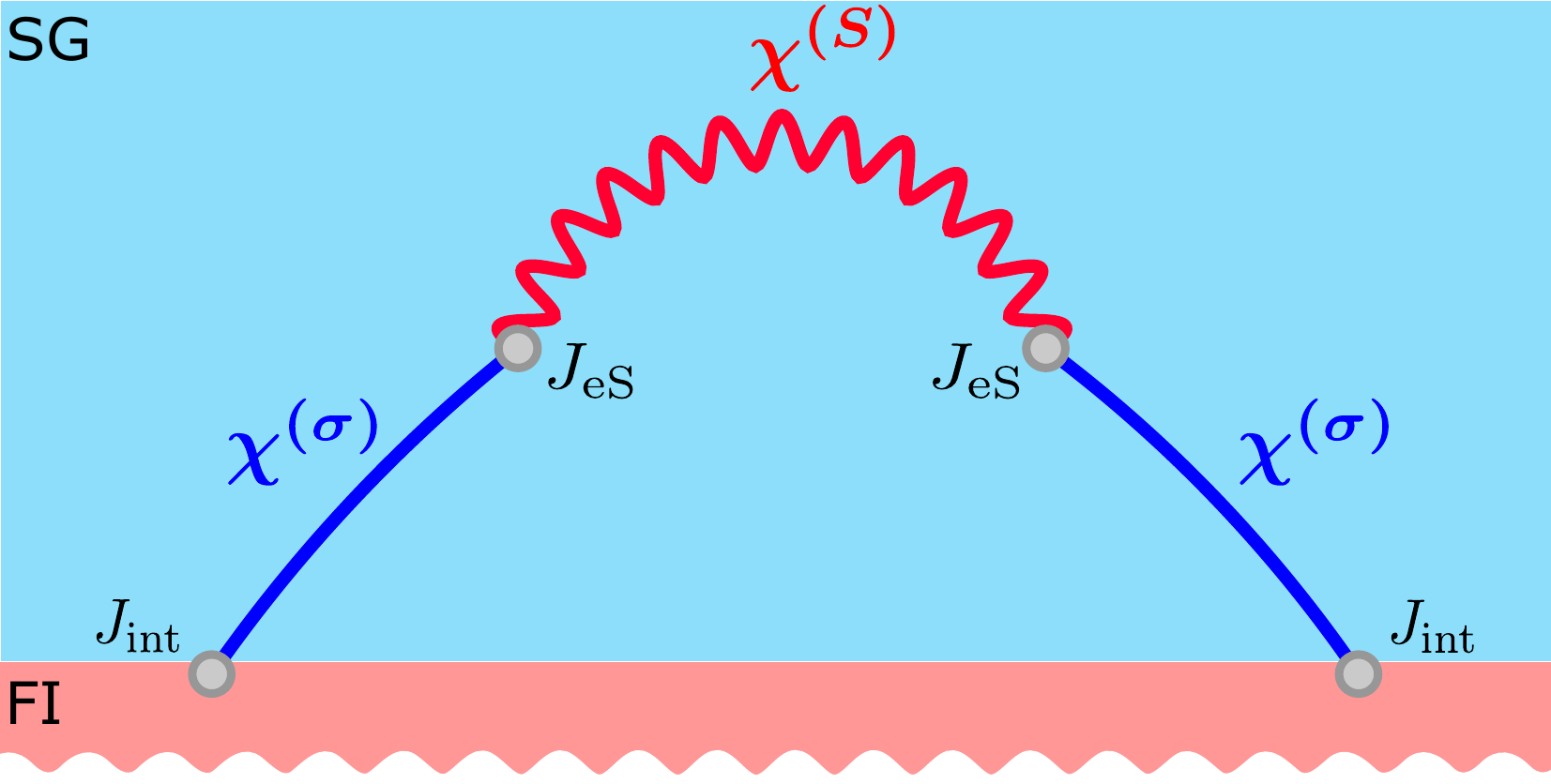}
  \end{center}
  \caption{Diagrammatic representation of the magnon self-energy giving the spin pumping signal. Here, $\chi^{(\sigma)} $ is the dynamic spin susceptibility of conduction-electron spins, whereas $\chi^{(S)}$ is that of impurity spins.}
  \label{fig:diagrammatic01}
\end{figure}

According to the linear-response formulation~\cite{Ohnuma14,Ohnuma15}, the additional magnon damping rate can be calculated from the corresponding magnon self-energy $\Sigma(\omega)$ whose process involves the spin transfer across the interface (Fig.~\ref{fig:diagrammatic01}). In the present situation, up to the lowest order with respect to $J_{\rm int}$, the self-energy is given by 
\begin{equation}
  \Sigma(\omega)= -\frac{J^2_{\rm int} N_{\rm int}}{\hbar^2 N_{\rm SG} N_{\rm FI}}
  \sum_{\bmq} \chi^{(\sigma)}_\bmq (\omega) J_{\rm eS} \chi^{(S)}_\bmq (\omega) J_{\rm eS} \chi^{(\sigma)}_\bmq (\omega), 
  \label{eq:Sigma01}
\end{equation}
where $N_{\rm int}$ is the number of the localized spins ${\bm \Omega}$ at the FI/SG interface, and $N_{\rm FI}$ is the number of lattice sites in the FI layer. In the above equation, $\chi^{(\sigma)}_\bmq (\omega)$ is the Fourier transform of the retarded susceptibility of the conduction-electron spin $\bmsig$, i.e., $\chi^{(\sigma)}_\bmq(t-t')= \ui \Theta(t-t') \bra [\sigma_\bmq^-(t),\sigma_{-\bmq}^+(t')] \ket$, where $\Theta(t)$ is the step function and we defined $O^\pm = O^x \pm \ui O^y$ for a vector operator ${\bm O}$. By contrast, $\chi^{(S)}_\bmq (\omega)$ is the Fourier transform of the retarded susceptibility of the impurity spin $\bmS$, i.e., $\chi^{(S)}_\bmq(t-t')= \ui \Theta(t-t') \bra [S_\bmq^-(t), S_{-\bmq}^+(t')] \ket$, 

Using the relation $\delta \alpha = - \omega_{\rm ac}^{-1}{\rm Im} \Sigma(\omega_{\rm ac})$~\cite{Ohnuma14}, the additional Gilbert damping constant $\delta \alpha$ is expressed as
\begin{equation}
  \delta \alpha \approx \frac{J^2_{\rm int} N_{\rm int}}{\hbar^2 N_{\rm FI}} 
  \left( \chi^{(\sigma)}_0 J_{\rm eS} \right)^2 \frac{1}{\omega_{\rm ac}} {\rm Im} \chi^{(S)}_{\rm loc} (\omega_{\rm ac}),
  \label{eq:Dalpha01}
\end{equation}
where we introduced a shorthand notation $\chi^{(\sigma)}_0= \chi^{(\sigma)}_{\bmq= {\bm 0}}(\omega=0)$, and $\chi^{(S)}_{\rm loc} (\omega)= N^{-1}_{\rm SG} \sum_\bmq \chi^{(S)}_{\bmq} (\omega)$ is the local susceptibility of impurity spins. In obtaining the above result, we made use of the fact that $\chi^{(S)}_{\rm loc} (\omega)$ is defined in a small $q$ region $q \lesssim 2 \pi/b$ where $b$ is the average distance of two magnetic impurities. In this small $q$ region, the conduction-electron spin susceptibility is approximated by the uniform and static component $\chi^{(\sigma)}_0$, where this quantity is pure real. 

Equation (\ref{eq:Dalpha01}) means that the additional Gilbert damping constant $\delta \alpha$ due to the spin pumping is proportional to the imaginary part of the dynamic spin susceptibility $\chi^{(S)}_{\rm loc} (\omega)$. In this expression, the strongest temperature dependence upon the SG transition results from the imaginary part of the dynamic spin susceptibility, ${\rm Im} [\chi^{(S)}_{\rm loc} (\omega)]$. This means that, as long as the temperature dependence is concerned, the part other than ${\rm Im} [\chi^{(S)}_{\rm loc} (\omega)]$ can be regarded as being temperature independent, and the temperature dependence is dominated by that of ${\rm Im} [\chi^{(S)}_{\rm loc} (\omega)]$. We adopt this approximation in the numerical calculation in Sec.~\ref{Sec:IV}. 

The quantity $\chi^{(S)}_{\rm loc} (\omega)$ is a correlation function between two impurity spins, and hence it can be evaluated using our knowledge on SGs. In the next section, we evaluate $\chi^{(S)}_{\rm loc} (\omega)$ using a dynamic theory of SGs~\cite{Sompolinsky82,Sompolinsky82b}.

\section{Dynamic spin susceptibility of impurity spins \label{Sec:III}}

In this section, by employing a dynamic theory of SGs~\cite{Sompolinsky82,Sompolinsky82b}, we sketch our procedure for calculating the dynamic spin susceptibility $\chi^{(S)}_{\rm loc} (\omega)$ appearing in the spin pumping signal [Eq.~(\ref{eq:Dalpha01})]. We emphasize that we use the dynamic model of the Heisenberg spin glasses developed by Sompolinsky and Zippelius~\cite{Sompolinsky82b}, which is constructed on top of a similar dynamic theory of Ising spin glasses~\cite{Sompolinsky82}. 

We first integrate out the conduction-electron degrees of freedom in the Hamiltonian for the SG layer ${\cal H}_{\rm SG}$. Then, ${\cal H}_{\rm SG}$ is transformed into the following form~\cite{Larkin71,Fischer79}: 
\begin{equation}
  \overline{\cal H}_{\rm SG} = \frac{1}{2} \sum_{i \neq j} J_{ij} \bmS(\bmr_i) \cdot \bmS(\bmr_j), 
  \label{eq:H_SG02}
\end{equation}
where $J_{ij}$ is nominally the RKKY interaction of the form $J_{ij}= J_{\rm eS} \cos(2 k_{\rm F} r_{ij} )/r_{ij}^3$ with $r_{ij}= |\bmr_i - \bmr_j|$. However, following the standard approach to the SG problem~\cite{Binder86}, we regards $J_{ij}$ as Gaussian random variables with zero mean and variance $[J_{ij}^2]_{\rm av}= J^2/N_S$, where $[\cdots]_{\rm av}$ means the random average over the distribution of $J_{ij}$, and $N_S$ is the number of impurity spins. 

Hamiltonian $\overline{\cal H}_{\rm SG}$ in Eq.~(\ref{eq:H_SG02}) is the same as the vector spin version~\cite{Kirkpatrick78} of the Sherrington-Kirkpatrick (SK) model~\cite{Sherrington75}, so that we employ the established dynamical approach. Following Sompolinsky and Zippelius~\cite{Sompolinsky82b}, we first replace $\overline{\cal H}_{\rm SG}$ with its soft-spin version: 
\begin{equation}
  \beta \widetilde{\cal H}_{\rm SG} = \frac{1}{2} \sum_\alpha \sum_{ij} (r_0 \delta_{ij}- \beta J_{ij}) S_i^\alpha S_j^\alpha
  - \beta \sum_i h_i^\alpha \cdot S_i^\alpha,
  \label{eq:H_SG03}
\end{equation}
where $\beta= 1/\kB T$ is the inverse temperature, and each component of the soft spin varies $-\infty < S_i^\alpha < \infty$, where the Greek superscript $\alpha= x, y, z$ specifies the direction in spin space. Here, we consider a paramagnet/SG transition by ignoring any tendency to ferromagnetic order, such that $r_0$ in Eq.~(\ref{eq:H_SG03}) is chosen to be a positive constant. Next, we introduce the Langevin dynamics 
\begin{equation}
  \Gamma_0^{-1} \partial_t S_i^\alpha = - \frac{\partial (\beta \widetilde{\cal H}_{\rm SG} )}{\partial S_i^\alpha}
  + \xi_i^\alpha(t),
    \label{eq:Langevin01}
\end{equation}
where $\Gamma_0$ is the bare relaxation rate. In the above equation, $\xi_i^\alpha(t)$ is a thermal noise represented by a Gaussian random variable with zero mean and variance
\begin{equation}
  \bra \xi_i^\alpha(t) \xi_j^{\alpha'} (t') \ket_{\bm \xi} = 2 \Gamma_0^{-1} \delta_{ij} \delta_{\alpha, \alpha'}\delta(t-t'),
  \label{eq:xixi01}
\end{equation}
where $\bra \cdots \ket_{\bm \xi}$ means the average over $\xi_i^\alpha$. In the following calculation, the response function 
\begin{equation}
  G_{ij}^{\alpha \alpha'}(t-t') = \left[ \frac{\partial \bra S_i^\alpha(t) \ket_{\bm \xi} }{\partial h_j^{\alpha'}(t')} \right]_{\rm av}, 
\end{equation}
plays an important role. This is because the spin susceptibility can be calculated by the relation
\begin{equation}
  \chi_{ij}^{\alpha \alpha'} (\omega) = \beta G_{ij} ^{\alpha \alpha'} (\omega),
  \label{eq:fluc-resp01}
\end{equation}
where $G_{ij} ^{\alpha \alpha'} (\omega)$ is the Fourier transform of $G_{ij} ^{\alpha \alpha'}(t-t')$. Besides, we need to define the correlation function 
\begin{equation}
  C_{ij} ^{\alpha \alpha'}(t-t') =  \left[ \bra S_i^\alpha(t) S_j^{\alpha'}(t') \ket_{\bm \xi} \right]_{\rm av}.  
\end{equation}

To proceed further, we precisely follow the procedure of Refs.~\cite{Sompolinsky82,Sompolinsky82b,Fischer84b}, which involves a lots of technical algebra. Since reviewing the details of Refs.~\cite{Sompolinsky82,Sompolinsky82b,Fischer84b} is beyond our scope, we leave it to the original paper and a famous textbook~\cite{Fischer-Hertz}, and we briefly sketch the derivation of the self-consistent dynamical equation in the mean-field limit. First, we rewrite Eq.~(\ref{eq:Langevin01}) in terms of a generating functional, and introduce an auxiliary field $\widehat{S}_i^\alpha$ as was done by Martin, Siggia, and Rose~\cite{MSR73}. Next, this generating functional is averaged over $J_{ij}$ without using replicas~\cite{Dominicis78}, which generates temporally nonlocal quartic interactions among $S_i^\alpha$ and $\widehat{S}_i^{\alpha'}$. Then, we introduce new auxiliary fields $Q_1^{\alpha \alpha'}, Q_2^{\alpha \alpha'}, Q_3^{\alpha \alpha'}$ and $Q_4^{\alpha \alpha'}$ to decouple the quartic terms, and we evaluate the functional integral by using the saddle-point approximation.

Following the above procedure, we obtain the new equation of motion for $S_i^\alpha$ containing the local self-interaction, which in the frequency space is written as 
\begin{eqnarray}
  - \frac{\ui \omega}{\Gamma_0} S^\alpha (\omega)
  &=& \sum_{\alpha'} \Big[ \big( -r_0 + \beta h^\alpha (\omega) \big) \delta_{\alpha \alpha'} \nonumber \\
  &&  + (\beta J)^2 G^{\alpha \alpha'}(\omega) \Big] S^{\alpha'}(\omega) 
   + \phi^\alpha(\omega), 
  \label{eq:Langevin02}
\end{eqnarray}
where $G^{\alpha \alpha'}(\omega)$ is the local response function $G_{ii}^{\alpha \alpha'}(\omega)$. Here and hereafter, the site index $i$ is discarded. In the above equation, $\phi^\alpha$ is a new noise field satisfying 
\begin{equation}
  \bra \phi^\alpha (\omega) \phi^{\alpha'} (\omega') \ket_{\bm \phi} 
  =  2 \pi \delta(\omega+ \omega')
  \left[ \frac{2 \delta_{\alpha \alpha'}}{\Gamma_0} + (\beta J)^2 C^{\alpha \alpha'}(\omega) \right], 
\end{equation}
where $C^{\alpha \alpha'}(\omega)$ is the Fourier transform of the local correlation function $C_{ii}^{\alpha \alpha'}(t-t')$, and $\bra \cdots \ket_{\bm \phi}$ is the average over $\phi^\alpha(\omega)$.

In the approach by Sompolinsky and Zippelius~\cite{Sompolinsky82b} the weak external-field limit was considered, such that {\it after the random average}, physical quantities are diagonal in spin space. Therefore, the local correlation function is separated as $C(t) = \mathfrak{q} + \Delta C(t)$, where $C^{\alpha \alpha'}(t)= C(t) \delta_{\alpha \alpha'} $ and $\mathfrak{q} = C(t)|_{t \to \infty}$. In parallel with this separation, the noise field $\phi^\alpha$ is divided into two parts, 
\begin{equation}
  \phi^\alpha(\omega)= f^\alpha(\omega)+ u^\alpha(\omega), 
\end{equation}
where the first term, $f^\alpha(\omega)$, satisfies 
\begin{equation}
  \bra f^\alpha(\omega) f^{\alpha'}(\omega') \ket =
  2 \pi \delta(\omega+ \omega') \delta_{\alpha \alpha'}
  \left[ \frac{2}{\Gamma_0} + (\beta J)^2 \Delta C(\omega) \right]
  \label{eq:ff01}
\end{equation}
with $\bra \cdots \ket$ being the usual thermal average over $f^\alpha(\omega)$, whereas the second term, $u^\alpha(\omega)$, satisfies
\begin{equation}
  [ u^\alpha(\omega) u^{\alpha'}(\omega') ]_\bmu
  = (2 \pi)^2 \delta(\omega+ \omega') \delta (\omega) \delta_{\alpha \alpha'} (\beta J)^2 \mathfrak{q},
  \label{eq:zz01}
\end{equation}
where $[ \cdots ]_\bmu$ is the average over $u^\alpha(\omega)$. 

From Eqs.~(\ref{eq:ff01}) and (\ref{eq:zz01}), we find that $f^\alpha(\omega)$ is a usual thermal fluctuation, whereas $u^\alpha(\omega)$ represents a frozen random field that breaks the ergodicity. Since $u^\alpha$ acts as a static random field, Sompolinsky and Zippelius introduced the ``unaveraged'' response function over $\bmu$: 
\begin{equation}
  g^{\alpha \alpha'}(\omega,\bmu) = \frac{\partial \bra S^\alpha(\omega) \ket }{\partial h^{\alpha'}(\omega)}, 
\end{equation}
where $g^{\alpha \alpha'}(\omega,\bmu)$ is related to $G^{\alpha \alpha'}(\omega)$ through 
\begin{equation}
  G^{\alpha \alpha'}(\omega)= [g^{\alpha \alpha'}(\omega,\bmu)]_\bmu. 
\end{equation}
Note that the average over $\bmu$ is separated as 
\begin{equation}
  [\cdots]_\bmu = [ [ \cdots ]_{\hat{\bm u}} ]_u,
\end{equation}
where $[ \cdots ]_{\hat{\bm u}}$ means the angular average over a unit vector $\hat{\bmu}= \bmu/u$, and from Eq.~(\ref{eq:zz01}) the average of a quantity $Q(u)$ over $u$ is given by 
\begin{equation}
  [Q(u)]_u =
  \int_{0}^\infty \frac{4 \pi u^2 du}{[2 \pi (\beta J)^2 \mathfrak{q}]^{3/2}}
  e^{-u^2/[2 (\beta J)^2 \mathfrak{q}]} Q(u).
\end{equation}

Now, we discuss the Dyson equation for $g^{\alpha \alpha'}(\omega,\bmu)$. In doing so, we introduce the matrix notation $\check{g}(\omega, \bmu)$, where the matrix element of $\check{g}(\omega, \bmu)$ equals $g^{\alpha \alpha'}(\omega,\bmu)$, i.e., 
\begin{equation}
  \check{g}(\omega, \bmu) = \Big( g^{\alpha \alpha'}(\omega,\bmu) \Big). 
\end{equation}
With this matrix notation the Dyson equation for $g(\omega,\bmu)$, which results from Eq.~(\ref{eq:Langevin02}) with static random field $\bmu$, is given by
\begin{equation}
  \check{g}(\omega,\bmu)^{-1} = \check{G}^{(0)}(\omega)^{-1}  - \check{\Sigma}(\omega,\bmu),
  \label{eq:Dyson01} 
\end{equation}
where the bare propagator is given by $\check{G}^{(0)}(\omega)^{-1}= [r_0- \ui \omega/\Gamma_0- (\beta J)^2 G(\omega) ]\check{\bm 1}$, and $\check{\Sigma}(\omega,\bmu)$ is the self-energy coming from the frozen $\bmu$-field. Since we are interested in the dynamic behavior of $\check{G}(\omega)$, we solve the Dyson equation by perturbation with respect to $\Delta \check{G}(\omega)= \check{G}(\omega)- \check{G}(0)$. We define $\check{\eta}(\omega,\bmu)= \ui \omega/\Gamma_0+ (\beta J)^2 \Delta \check{G}(\omega)- \Delta \check{\Sigma}(\omega,\bmu)$, where $\Delta \check{\Sigma}(\omega,\bmu)= \check{\Sigma}(\omega,\bmu)- \check{\Sigma}(0,\bmu)$, and rewrite the Dyson equation as $\check{g}(\omega,\bmu)^{-1}= \check{g}(0,\bmu)^{-1}- \check{\eta}(\omega,\bmu)$. To proceed further, we disregard $\Delta \check{\Sigma}(\omega,\bmu)$ as it brings only a small change~\cite{Fischer84a}. This allows us to regard $\check{\eta}(\omega,\bmu)$ as $\bmu$-independent, and the matrix $\check{\eta}$ becomes diagonal in spin space, i.e., $\check{\eta}(\omega)= \eta(\omega) \check{\bm 1}$. Then, after expanding the Dyson equation up to $\check{\eta}^2$ and averaging the result over $\bmu$, we obtain a quadratic equation for $\Delta \check{G}(\omega)= \Delta {G}(\omega) \check{\bm 1}$: 
\begin{eqnarray}
  \Delta {G}(\omega) \left( 1- (\beta J)^2 [\check{g}_0^2]_\bmu \right) 
  &=& \left( \ui \frac{\omega}{\Gamma_0}
  + (\beta J)^2 \Delta {G}(\omega) \right)^2 [\check{g}_0^3]_\bmu \nonumber \\ 
  && +\ui \frac{\omega}{\Gamma_0} [\check{g}_0^2]_\bmu ,
  \label{eq:Dyson02}   
\end{eqnarray}
where we introduced the shorthand notation $[\check{g}_0^n]_\bmu= [\check{g}(0,\bmu)^n]_\bmu$. Note that $[\check{g}_0^n]_\bmu$ is diagnoal in spin space {\it after the average over} $\bmu$, such that it appears in Eq.~(\ref{eq:Dyson02}) as a c-number. Equation (\ref{eq:Dyson02}) can be solved for $\Delta G(\omega)$. After using the relation in Eq.~(\ref{eq:fluc-resp01}), we obtain 
\begin{equation}
  \chi^{(S)}_{\rm loc}(\omega) = \chi^{(S)}_{\rm loc}(0)+
  \frac{\widetilde{T}/3}{2 [\check{g}_0^3]_\bmu}
  \left( {\cal D} - \sqrt{{\cal D}^2 - \ui \frac{4 \omega}{\Gamma_0} [\check{g}_0^2]_\bmu[\check{g}_0^3]_\bmu} 
  \right),
  \label{eq:chi_omega01}
\end{equation}
where ${\cal D} = (\widetilde{T}/3)^2- [\check{g}_0^2]_\bmu - 2 \ui ({\omega}/\Gamma_0)[\check{g}_0^3]_\bmu $, $\widetilde{T}= T/T_{g}$. Note that, in contrast to the Ising spin glass~\cite{Sherrington75}, the prenset Heisenberg spin glass has a transition temperature $T_{g}= J/3 \kB$~\cite{Kirkpatrick78}. 

In the following calculation, we take the mean-field approximation of the classical Heisenberg spins under a frozen $\bmu$ field. We use $\bra S^\alpha \ket= \bra S \ket \hat{u}^\alpha$ and $\bra S^\alpha S^{\alpha'} \ket= \delta_{\alpha \alpha'}/3$, where $\bra S \ket= \coth(u)- 1/u$ is the Langevin function. Then, recalling that $[\bra S^\alpha S^{\alpha'} \ket]_\bmu= [\bra S \ket^2]_u [\hat{u}^\alpha \hat{u}^{\alpha'}]_{\hat{\bmu}}= [\bra S \ket^2]_u \delta_{\alpha \alpha'}/3$, the spin glass order parameter $\mathfrak{q}$ can be calculated from 
\begin{equation}
  \mathfrak{q} = \frac{1}{3} [\bra S \ket^2]_u , 
\end{equation}
where the quantity $[\bra S \ket^{2s} ]_u$ is given by
\begin{eqnarray}
  [\bra S \ket^{2s}]_u &=& \int_{0}^\infty \frac{4 \pi u^2 du}{[2 \pi (\beta J)^2 \mathfrak{q}]^{3/2}} 
  e^{-u^2/[2 (\beta J)^2 \mathfrak{q}]} \nonumber \\
  && \qquad \times  \left( \coth(u)- \frac{1}{u} \right)^{2s}. 
\end{eqnarray}
In a similar manner, we use $[\check{g}_0]_\bmu= (1- 3\mathfrak{q})/3 $, $[\check{g}_0^2]_\bmu= (1- 6\mathfrak{q}+ 3[\bra S \ket^4]_u)/9 $, and $[\check{g}_0^3]_\bmu= (1- 9 \mathfrak{q}+ 9[\bra S \ket^4]_u - 3[\bra S \ket^6]_u)/27 $. At the SG transition, the Almeida-Thouless condition~\cite{Almeida78} holds, which in the present notation takes the form 
\begin{equation}
  [\check{g}_0^2]_\bmu= \left( {\widetilde{T}}/ {3} \right)^2.
  \label{eq:AT01}
\end{equation}
Then, following Sompolinsky and Zippelius~\cite{Sompolinsky82} we assume that Eq.~(\ref{eq:AT01}) holds not only at $T_g$ but also below $T_g$, meaning that the SG phase is characterized by the marginal instability condition.

\begin{figure}[t] 
  \begin{center}
    \includegraphics[width=8.0cm]{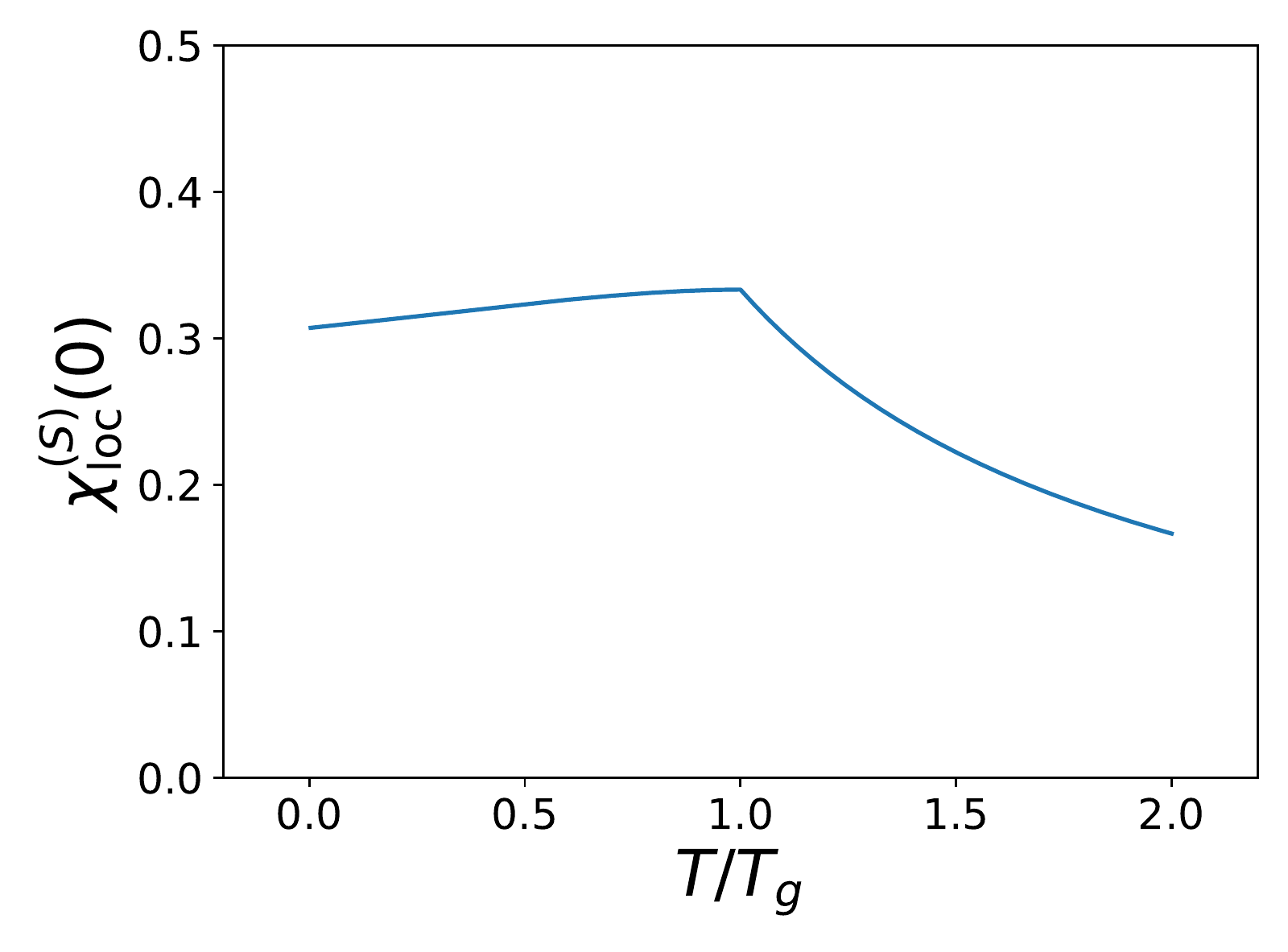}
  \end{center}
  \caption{Static spin susceptibility $\chi^{(S)}_{\rm loc} (0)$ as a function of temperature calculated using Eq.~(\ref{eq:Fischer-relation01}).} 
  \label{fig:Rechi01}
\end{figure}

\section{Results for spin pumping into spin glass materials \label{Sec:IV}} 

\begin{figure}[t] 
  \begin{center}
    \includegraphics[width=8.8cm]{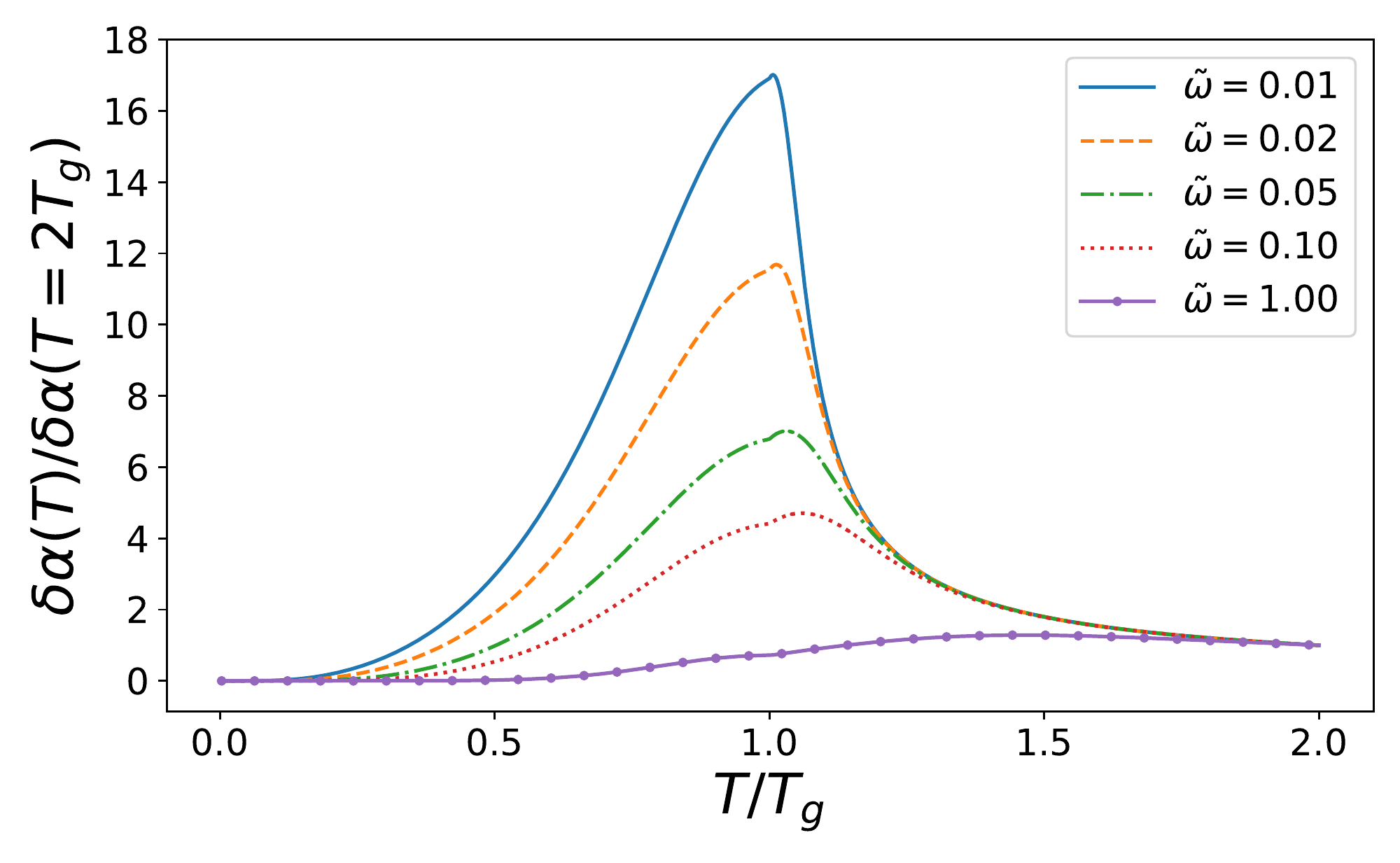}    
  \end{center}
  \caption{Temperature dependence of the additional Gilbert damping constant [Eq.~(\ref{eq:Dalpha01})] in a SG/FI bilayer (Fig.~\ref{fig:schematic01}), calculated for several values of $\widetilde{\omega}_{\rm ac}= \omega_{\rm ac}/\Gamma_0$.}   
  \label{fig:Imchi01}
\end{figure}

In this section, using the formalism developed in the previous two sections, we calculate temperature dependence of the spin pumping signal. The key equation in the present argument is Eq.~(\ref{eq:Dalpha01}), which relates the spin pumping with the dynamic spin susceptibility of SG materials.

Before presenting our results for the spin pumping that is intimately related to the {\it dynamic} spin susceptibility $\chi^{(S)}_{\rm loc} (\omega)$, it is instructive to examine the {\it static} spin susceptibility $\chi^{(S)}_{\rm loc} (0)$. This quantity can be calculated using~\cite{Fischer76}: 
\begin{equation}
  \chi^{(S)}_{\rm loc}(0)= \frac{1}{3\widetilde{T}}(1- 3 \mathfrak{q}), 
  \label{eq:Fischer-relation01}
\end{equation}
where the additional factor $3$ follows from the relation $[g^{\alpha \alpha'}]_\bmu= \delta_{\alpha \alpha'} (1- 3\mathfrak{q})/3 $. Figure~\ref{fig:Rechi01} shows the static spin susceptibility $\chi^{(S)}_{\rm loc} (0)$ as a function of temperature, calculated from Eq.~(\ref{eq:Fischer-relation01}). The result reproduces the well-known cusp structure at $T_g$~\cite{Sherrington75,Kirkpatrick78}. 

Now we discuss the spin pumping into a SG material. In discussing the spin pumping signal in the present case, the important parameter is the ratio of the microwave angular frequency $\omega_{\rm ac}$ to the relaxation rate of localized spins $\Gamma_0$, i.e., $\widetilde{\omega}_{\rm ac} \equiv \omega_{\rm ac}/\Gamma_0$. Since the magnitude of $\omega_{\rm ac}$ under the FMR condition is of the order of $60$~GHz, the parameter $\widetilde{\omega}_{\rm ac}$ is determined by a material parameter $\Gamma_0$. The spin relaxation time in a prototypical SG material Cu:Mn is reported to be of the order of picosecond (corresponding to $\Gamma_0 \sim 10^3$~GHz)~\cite{Mezei82}, and hence the parameter $\widetilde{\omega}_{\rm ac}$ under the FMR condition is estimated to be $\widetilde{\omega}_{\rm ac} \sim 0.1$. Note that the previous calculation of the dynamic spin susceptibility~\cite{Fischer84a} was done for very low frequencies $\widetilde{\omega}_{\rm ac} \lesssim 10^{-4}$, which is far out of the FMR condition. 

Figure \ref{fig:Imchi01} shows temperature dependence of the spin pumping signal, calculated from Eqs.~(\ref{eq:Dalpha01}) and (\ref{eq:chi_omega01}). First, we see that a clear peak structure appears at $T_g$. Second, upon the increase of the parameter $\widetilde{\omega}_{\rm ac}$, the height of the peak is reduced. This is because the peak structure originates from the critical slowing down of spins that develops on the verge of the spin freezing~\cite{Fischer-Hertz}, so that the effects of the slowing down are more prominent when the paramagnetic state has a more rapid dynamics (i.e., larger $\Gamma_0$) in comparison to the spin frozen state. 

As shown in Fig.~\ref{fig:Imchi01}, the peak structure is visible for a parameter region $\widetilde{\omega}_{\rm ac} \lesssim 0.1$. Since the parameter $\widetilde{\omega}_{\rm ac}$ for a prototypical SG material Cu:Mn is estimated about $\widetilde{\omega}_{\rm ac} \sim 0.1$~\cite{Mezei82} as mentioned above, we expect that we can observe a peak structure in the spin pumping signal near $T_g$. Therefore, we propose a spin pumping experiment for a Cu:Mn/YIG system in order to test our theoretical prediction. Frequency ($\omega_{\rm ac}$) dependence of the peak mentioned above, namely, the lower the frequency $\omega_{\rm ac}$, the higher the peak, is a key to identify the predicted signal.

\section{Discussion and Conclusion \label{Sec:V}}
The main result of this paper is the theoretical prediction that the spin pumping into a SG material, whose signal is proportional to the additional Gilbert damping constant $\delta \alpha$, can be enhanced near the SG transition. The physics behind this enhancement is explained in the following way. First, a SG material exhibits the critical slowing down upon the spin freezing~\cite{Fischer-Hertz}, meaning that the spin relaxation rate of the SG material is reduced. Then, recalling that the spin pumping represents an additional damping of magnons in the spin injecting magnet, this type of enhancement in the spin pumping signal can be interpreted as a kind of the inverse of the motional narrowing~\cite{Kubo-text,Slichter-text} as discussed in Ref.~\cite{Taira18} (see Sec.~VI therein). It means that a reduction of the spin relaxation rate in the spin sink material results in a broadening of the magnon damping in the adjacent magnet, leading to the enhancement of the spin pumping.

The spin pumping has an advantage that it can measure the dynamic spin susceptibility of a {\it thin} film sample. So far, this fact has been applied to the spin pumping into ferromagnets~\cite{Ohnuma14,Khodadadi17}, antiferromagnets~\cite{Frangou16,Qiu16}, and superconductors~\cite{Inoue17,Yao18,Umeda18}. Extending the same idea to SGs within the linear-response approach, we formulated the spin pumping into a SG material, and shown that the signal is expressed by using the local spin susceptibility of the SG material [Eq.~(\ref{eq:Dalpha01})]. Moreover, the spin pumping into a SG material is predicted to exhibit a characteristic peak around $T_g$ (Fig.~\ref{fig:Imchi01}).  

The height of the predicted peak in the spin pumping signal is controlled by the ratio of the microwave angular frequency to the relaxation rate of impurity spins, i.e., $\widetilde{\omega}_{\rm ac}= \omega_{\rm ac}/\Gamma_0$. That is, a smaller $\widetilde{\omega}_{\rm ac}$ is better for an experimental detection of the peak. Conversely, it means that if the parameter $\widetilde{\omega}_{\rm ac}$ is too large, the peak is not visible. To test our theoretical prediction, we hope a future spin pumping experiment using an insulating magnet with, e.g., Cu:Mn/YIG structure, since use of an insulating magnet instead of a metallic magnet makes the spin pumping signal more visible owing to the smallness of the intrinsic Gilbert damping term $\alpha_0$.

Before conclusion, we briefly comment on the previous experiment of the spin pumping into a SG material using Ag$_{90}$Mn$_{10}$/Ni$_{81}$Fe$_{19}$~\cite{Iguchi11}. In that experiment, the SG transition temperature is estimated to be $T_g = 25$~K from the cusp in the susceptibility data of a {\it thick} Ag$_{90}$Mn$_{10}$ film. Note that this {\it thick} film is different from the {\it thin} film used for the spin pumping experiment. In the spin pumping experiment, a weak temperature dependence of the signal was measured around $T_g$, but no pronounced peak expected from the theoretical calculation (Fig.~\ref{fig:Imchi01}) can be seen. The reason could be either i) the important parameter $\widetilde{\omega}_{\rm ac} = \omega_{\rm ac}/\Gamma_0$ is too large in Ag$_{90}$Mn$_{10}$ for the enhanced spin pumping to be detected (see Fig.~\ref{fig:Imchi01}), or ii) the SG transition temperature of the {\it thin} film sample is much lower than $25$~K, since the {\it thin} film may have lower $T_g$ than the {\it thick} film used to determine $T_g = 25$~K.

To conclude, we have theoretically examined the spin pumping into a SG material. We have shown that the temperature dependence exhibits a characteristic peak near the SG transition, whose height is controlled by the dimensionless angular frequency $\widetilde{\omega}_{\rm ac}= \omega_{\rm ac}/\Gamma_0$. This is a consequence of the critical slowing down of spin fluctuations upon the spin freezing. Since the spin pumping has an advantage of being able to measure the spin dynamics of a {\it thin} film sample~\cite{Inoue17}, we hope that the present theory stimulates further experiments of the spin pumping into SG materials. 

\acknowledgments 
This work was financially supported by JSPS KAKENHI Grant Number 19K05253.




\end{document}